\documentclass[aps,prx,10pt,twocolumn,superscriptaddress]{revtex4-1}

\usepackage{times}
\usepackage{changes}
\usepackage{graphicx}
\usepackage{amssymb}
\usepackage{epstopdf}
\usepackage{amsmath,bm}
\usepackage{color}
\usepackage{hyperref}
\usepackage{bbold}
\usepackage{mathtools}
\usepackage{enumitem}
\usepackage[acronym]{glossaries}
\usepackage{physics}
\usepackage{booktabs}
\makeatletter
\renewcommand{\arraystretch}{1.2}

\@ifundefined{textcolor}{}
{
	\definecolor{BLACK}{gray}{0}
	\definecolor{WHITE}{gray}{1}
	\definecolor{RED}{rgb}{1,0,0}
	\definecolor{GREEN}{rgb}{0,1,0}
	\definecolor{darkgreen}{rgb}{.1,.6,.1}
	\definecolor{BLUE}{rgb}{0,0,1}
	\definecolor{CYAN}{cmyk}{1,0,0,0}
	\definecolor{MAGENTA}{cmyk}{0,1,0,0}
	\definecolor{YELLOW}{cmyk}{0,0,1,0}
}

\graphicspath{{Images/}}

\hypersetup{
	pdfpagemode={UseOutlines},
	bookmarksopen,
	pdfstartview={FitH},
	colorlinks,
	linkcolor={black},
	citecolor={blue},
	urlcolor={blue}}

\usepackage{hyperref}
\hypersetup{
	colorlinks=true,
	linkcolor=black,          
	citecolor=blue,           
	filecolor=magenta,        
	urlcolor=cyan
}

\makeatother	

\begin{document}
	
	\title{Bridging the gap between classical and quantum many-body information dynamics}
	
	\author{Andrea Pizzi}
	\email[Corresponding author, ]{ap2076@cam.ac.uk}
	\affiliation{Cavendish Laboratory, University of Cambridge, Cambridge CB3 0HE, United Kingdom}
	\author{Daniel Malz}
	\affiliation{Max-Planck-Institute of Quantum Optics, Hans-Kopfermann-Str.~1, 85748 Garching, Germany}
	\affiliation{Munich Center for Quantum Science and Technology (MCQST), 80799 Munich, Germany}
	\author{Andreas Nunnenkamp}
	\affiliation{Faculty of Physics, University of Vienna, Boltzmanngasse 5, 1190 Vienna, Austria}
	\author{Johannes Knolle}
	\affiliation{Department of Physics, Technische Universit{\"a}t M{\"u}nchen, James-Franck-Stra{\ss}e 1, 85748 Garching, Germany}
	\affiliation{Munich Center for Quantum Science and Technology (MCQST), 80799 Munich, Germany}
	\affiliation{Blackett Laboratory, Imperial College London, London SW7 2AZ, United Kingdom}
	
	\begin{abstract}
		The fundamental question of how information spreads in closed quantum many-body systems is often addressed through the lens of the bipartite entanglement entropy, a quantity that describes correlations in a comprehensive (nonlocal) way. Among the most striking features of the entanglement entropy are its unbounded linear growth in the thermodynamic limit, its asymptotic extensivity in finite-size systems, and the possibility of measurement-induced phase transitions, all of which have no obvious classical counterpart. Here, we show how these key qualitative features emerge naturally also in classical information spreading, as long as one treats the classical many-body problem on par with the quantum one, that is, by explicitly accounting for the exponentially large classical probability distribution. Our analysis is supported by extensive numerics on prototypical cellular automata and Hamiltonian systems, for which we focus on the classical mutual information and also introduce a `classical entanglement entropy'. Our study sheds light on the nature of information spreading in classical and quantum systems, and opens new avenues for quantum-inspired classical approaches across physics, information theory, and statistics.
	\end{abstract}
	
	\maketitle
	
	\section{Introduction}
	\label{sec. Introduction}
	Many-body quantum systems display a huge variety of physical phenomena and may carry a vast amount of information in the exponentially many components of their wavefunction.
	Characterizing this information has become a major goal of modern quantum science, of prime relevance for quantum computing~\cite{nielsen2002quantum} and quantum simulation~\cite{georgescu2014quantum, altman2021quantum}.
	In recent years the study of closed many-body quantum systems has flourished in particular in the nonequilibrium regime, with key questions revolving around the dynamics of equilibration and information spreading~\cite{polkovnikov2011colloquium, gogolin2016equilibration}.
	These have become increasingly relevant in light of the experimental advances with nonequilibrium many-body systems kept in almost isolated conditions, e.g., in Rydberg atom arrays~\cite{browaeys2020many} or cold atoms in optical lattices~\cite{gross2017quantum}.
	
	One of the most prominent tools that has emerged from this field is the bipartite entanglement entropy (EE). As it can account for correlations in a comprehensive, nonlocal, multi-point way, the EE has been extensively adopted to monitor the dynamical entangling of the system's parts in pure quantum systems~\cite{calabrese2005evolution, de2006entanglement, amico2008entanglement, calabrese2009entanglement, bardarson2012unbounded, kim2013ballistic, ho2017entanglement, nahum2017quantum, nahum2018operator, khemani2018operator, von2018operator, gopalakrishnan2018hydrodynamics, alba2018entanglement, abanin2019colloquium, zhou2019emergent, bertini2019entanglement, ippoliti2022fractal}. Among the paradigms that have been unearthed are the linear and logarithmic unbounded growth of the EE in the thermodynamic limit for generic many-body systems~\cite{abanin2019colloquium} and in many-body localized (MBL) ones~\cite{bardarson2012unbounded,serbyn2013universal}, respectively, and its saturation to an extensive value for finite-size systems at long times. More recently, it has been shown that this extensivity allows for novel measurement-induced phase transitions (MIPT), whereby the rate of local random measurements determines whether the asymptotic EE scales proportional to the volume of the considered system's parts (`volume-law scaling'), or to the area of the boundary between them (`area-law scaling'). Since its discovery three years ago~\cite{skinner2019measurement}, the MIPT has received a remarkable amount of interest~\cite{li2019measurement, bao2020theory, choi2020quantum, tang2020measurement, jian2020measurement, zabalo2020critical, tang2020measurement, nahum2021measurement, ippoliti2021entanglement, block2022measurement}.
	
	\renewcommand{\arraystretch}{1.1}
	\begin{table*}[th]
		\centering
		\begin{tabular}{@{}cccc@{}}
			\toprule[1pt]
			& \textbf{Classical} & \textbf{Quantum} & Reference \\ 
			\midrule[0.5pt]
			Space & phase space & Hilbert space \\
			$N$-Particle space size & $\sim e^{\mathcal{O}(N)}$ & $\sim e^{\mathcal{O}(N)}$ \\
			State of the system & probability distribution $p$ & wavefunction $\ket{\psi}$ \\
			Bipartite (c)EE $S_e$ & Eq.~\eqref{eq. cEE} & Eq.~\eqref{eq. EE} & Sec.~\ref{sec. theory} \\
			$S_e = 0$ if... & $p_{A,B} = p_A p_B$ & $\ket{\psi} = \ket{\psi_A} \ket{\psi_B}$ & Sec.~\ref{sec. theory} \\
			Few-point observables become$^\dagger$ & thermal & thermal & Sec.~\ref{sec. thermalization} \\
			The actual MI can be$^\dagger$ & volume law & volume law & Secs.~\ref{sec. CA}, \ref{sec. Hamiltonian dynamics} \\
			The MI of a thermal ensemble is$^\dagger$ & area law & area law & Ref.~\cite{wolf2008area} \\
			The probability/wavefunction becomes$^\dagger$ & effectively randomized & effectively randomized & Sec.~\ref{sec. thermalization}\\
			Origin of effective randomization$^\dagger$ & chaos and incompressibility & chaotic spectrum and dephasing & Sec.~\ref{sec. thermalization}\\
			The effective randomization underlines$^\dagger$ & volume law cEE & volume law EE & Ref.~\cite{page1993average}\\
			For $N \to \infty $ the EE shows$^\dagger$ & unbounded linear growth & unbounded linear growth & Figs.~\ref{cEEfig2}, \ref{cEEfig6}\\
			Site $j$ is measured according to... & the marginal probability $p_j$ & the reduced density matrix $\rho_j$ & Sec.~\ref{sec. measurements}\\
			Right after a measurement, site $j$ is... & `factored out', $p = p_j p_{\setminus j}$ & `factored out', $\rho = \rho_j \rho_{\setminus j}$ & Sec.~\ref{sec. measurements}\\
			Information spreading vs measurements can lead to & MIPT & MIPT & Sec.~\ref{sec. measurements}\\
			Computing the (c)EE classically requires & exponentially large resources & exponentially large resources & Sec.~\ref{sec. cl vs qu} \\
			Extracting the (c)EE experimentally requires & exponentially many runs & exponentially many runs & Sec.~\ref{sec. cl vs qu} \\
			\bottomrule[1pt]
		\end{tabular}
		\caption{\textbf{Bridging the classical-quantum gap in many-body information spreading.} We summarize the key differences and, mostly, close analogies between classical and quantum information spreading. The setting we refer to is one in which initially local classical or quantum fluctuations become nonlocal under some time-reversible and local dynamics. Disclaimer: the lines marked with $^\dagger$ are valid under the general circumstances specified in the main text. The last column points to suitable Sections, Figures, or References in which each comparison can be best appreciated.}
		\label{tab. analogies}
	\end{table*}
	
	An obvious and conceptually fundamental question is: to what extent are these distinctive and celebrated many-body features of quantum information spreading purely quantum? One of the reasons why they have become well-appreciated in the quantum world is that state-of-the-art numerical techniques can approximate~\cite{schollwock2011density} or even exactly describe~\cite{weisse2008exact, weinberg2017quspin} the exponentially large many-body wavefunction. In contrast, the probability distribution of classical nonequilibrium many-body systems, while also being exponentially large, is normally not explicitly accounted for. Rather, classical information spreading is mostly investigated in terms of spatio-temporal correlations, transport properties, and the spreading of perturbations~\cite{vastano1988information, lepri1996chronotopic, giacomelli2000convective, das2018light, bilitewski2018temperature, khemani2018velocity, kumar2020transport, bilitewski2021classical,liu2021butterfly, mcroberts2022anomalous, deger2022arresting}, all of which are based on the study of few-body observables that can be effectively estimated à la Monte Carlo as averages over a handful of trajectories. According to the paradigm of thermalization, the expected value of these few-point observables reaches, at long times and under general circumstances, the value predicted by a thermal ensemble $\propto e^{-\beta H}$ at suitable temperature $\beta^{-1}$~\cite{kardar2007statistical}. But for instance, the bipartite mutual information (MI), that is often regarded as the natural classical analogue of the bipartite EE, is many-body in nature, requires the knowledge of the system's full probability distribution, and at long times does not obviously reach the value predicted by a thermal ensemble.
	
	Here, we argue that this mismatch in the description of quantum wavefunctions and classical probability distributions has led to a gap between our understanding of classical and quantum information spreading. We bridge this gap by (i) accounting for the dynamics of the full many-body probability distribution, thus treating the classical many-body problem on par with the quantum one, and (ii) focusing on many-body information measures akin to quantum EE, such as the classical MI and the `classical EE' (cEE), a complementary measure for classical non-separability that we introduce. We find that the qualitative features of quantum and classical information spreading are remarkably similar, and show the first instances of asymptotic extensivity, unbounded linear growth, and MIPT of MI and cEE in a classical setting. The analogies between classical and quantum information spreading are summarized in Table~\ref{tab. analogies}.
	
	The rest of the paper is structured as follows. In Section~\ref{sec. theory} we introduce the main notation and our definition of the cEE. In Section~\ref{sec. thermalization} we review the idea of thermalization following a quench to remark on the importance of retaining information on the full and exponentially large state of the system, both in the classical and quantum cases. In Section~\ref{sec. CA} we show how the main features of quantum information spreading naturally emerge in classical cellular automata: the bipartite MI and cEE grow linearly in time until saturating to an extensive value. Emphasis is put on the role that time reversibility plays in these effects. Upon interleaving the automaton dynamics with local measurements, we find and analyse a classical MIPT in Section~\ref{sec. measurements}. The discussion moves then to continuous Hamiltonian dynamics in Section~\ref{sec. Hamiltonian dynamics}, where with analytical arguments and numerics we show the asymptotic MI to be generally extensive, reducing to area-law only for initial conditions at an effective infinite temperature. We highlight a key difference between the MI and the cEE, namely that the cEE remains asymptotically extensive even at infinite temperature, thus further narrowing the classical-quantum gap. In Section~\ref{sec. cl vs qu} we argue that, beyond sharing the same phenomenology, classical and quantum many-body information spreading also require remarkably similar experimental and computational protocols to be observed, involving in both cases either exponentially-large resources or exponentially many runs. We conclude in Section~\ref{sec. discussion} with a discussion of the results and an outlook.
	
	\section{Notation and definitions}
	\label{sec. theory}
	Consider a bipartite system $(A,B)$ consisting of either classical or quantum degrees of freedom. We are interested in quantifying the amount of information that the parts $A$ and $B$ carry on one another. The system is described by a probability distribution $p$ and by a density matrix $\rho$ in the classical and quantum cases, respectively. While we are here ultimately interested in the study of classical systems, to most easily appreciate the classical-quantum connections we will try to develop the formalism in clear analogy with the quantum one. Whether we are talking about a classical or a quantum system should be clear from context and notation.
	
	The part $A$ itself is described by the marginal probability distribution $p_A = \Tr_B p = \sum_B p_{A,B}$ or by the reduced density matrix $\rho_A = \Tr_B \rho$, with $\Tr_B$ the partial trace over $B$. The classical entropy reads $S = - \sum_{A,B} p_{A,B} \log p_{A,B}$, whereas the quantum (von Neumann) entropy yields $S = - \Tr \left[\rho \log \rho\right]$. Similarly, the marginal entropy associated with $A$ reads $S_A = -\sum_A p_A \log p_A$ or $S_A = -\Tr_A \left[\rho_A \log \rho_A \right]$, and analogously for $S_B$.
	
	A central object in information theory, statistics, and statistical physics is the MI, quantifying by how much our ignorance on $B$ is reduced when observing $A$~\cite{cover2006elements}. Among its possible representations, one that holds for both the classical and quantum cases is
	\begin{equation}
	I_{A;B} = S_A + S_B - S.
	\label{eq. MI}
	\end{equation}
	The MI is non-negative, and vanishes if and only if the state is separable, that is, when $p_{A,B} = p_A p_B$ or $\rho = \rho_A \otimes \rho_B$. Thus, the MI unambiguously diagnoses any statistical interdependence between $A$ and $B$.
	
	Generally, the marginal entropies $S_A$ and $S_B$ are not representative of the degree of interdependence of $A$ and $B$, but just of the degree of uncertainty on $A$ and $B$ themselves: $S_A$ and $S_B$ can be positive while $I_{A;B}$ vanishes. Still, special cases exist in which $S_A$ and $S_B$ are proportional to the MI. Most notably, for pure quantum states ($\rho = \ket{\psi} \bra{\psi}$) one has $S = 0$, and thus $S_A = S_B = \frac{1}{2} I(A,B)$. Indeed, in this case $S_A$ is known as EE. Since $S_A$ is defined for generic mixed states, but can only be called EE when used on pure states, it is worth with a bit of redundancy to introduce a new symbol $S_e$ for the EE, whose explicit expression reads
	\begin{equation}
	S_e = - \Tr_A\left( \rho_A \log \rho_A \right),
	\quad
	\rho_A = \Tr_B\left(\ket{\psi} \bra{\psi}\right).
	\label{eq. EE}
	\end{equation}
	
	The MI is a natural classical counterpart of the EE. Yet, seeking to make the classical-quantum analogy even more direct, we make an extra step and propose an alternative analogue of the EE, the cEE, whose definition is closely inspired by the quantum one. As it will turn out, the cEE complements the MI, showing behaviours analogue to the quantum EE even when the MI fails to do so. To define it, we first introduce the `classical reduced density matrix' $\rho_A$ as the matrix with entries
	\begin{equation}
	\left(\rho_A\right)_{A',A''} = \sum_{B} \sqrt{p_{A',B}p_{A'',B}}.
	\end{equation}
	The diagonal elements of $\rho_A$ yield the marginal probability distribution, $\left(\rho_A\right)_{A,A} = p_A$, and so $\Tr \rho_A = 1$. Since $\rho_A$ is positive semi-definite
	\footnote{The reduced density matrix can be expressed as $\rho_A = \left(P^{\left(1/2\right)}\right)^T P^{\left(1/2\right)}$, where $P^{\left(1/2\right)}$ is a `squared probability matrix' with entries $P^{\left(1/2\right)}_{A,B} = \sqrt{p_{A,B}}$. Given a vector $z$, we have that $z^T \rho_A z = z^T \left(P^{\left(1/2\right)}\right)^T P^{\left(1/2\right)} z$. Calling $y = P^{\left(1/2\right)} z$, we therefore have that $z^T \rho_A z = y^T y \ge 0$, that is, that $\rho_A$ is positive semi-definite.}
	and $\Tr \rho_A = 1$, its eigenvalues $\{\lambda_n\}$ represent a valid (positive and normalized) probability distribution, which we use to define the cEE as
	\begin{equation}
	S_e = - \Tr \left[\rho_A \log \rho_A \right]
	= - \sum_{n} \lambda_n \log \lambda_n.
	\label{eq. cEE}
	\end{equation}
	Note that the cEE $S_e$ is nothing but the quantum EE associated to the `classical wavefunction'~\cite{wetterich2017fermions, wetterich2020probabilistic} $\ket{\psi}$ with components
	\begin{equation}
	\braket{A,B}{\psi} \coloneqq \sqrt{p_{A,B}},
	\label{eq. classical wavefunction}
	\end{equation}
	which is correctly normalized, $\braket{\psi}{\psi} = \sum_{A,B} p_{A,B} = 1$, and which lives in a fictitious Hilbert space generated by considering the configurations of $(A,B)$ as a basis $\{\ket{A,B}\}$. By construction, many properties of the cEE thus directly follow from those of its quantum counterpart. Most importantly, $S_e \ge 0$, and $S_e = 0$ if and only if $I_{A;B} = 0$, which makes the cEE a good witness of statistical dependence between $A$ and $B$, in contrast to the marginal entropy $S_A$ (the latter is computed on the diagonal of $\rho_A$ rather than on its eigenvalues). Moreover, we note that, while for pure quantum states the MI and the EE coincide (up to a factor 2), the classical MI and cEE are generally different. Indeed, while they both quantify statistical dependence between $A$ and $B$, the cEE features scaling behaviours that the MI lacks, and that are crucial to connect to quantum information spreading.
	
	\begin{figure*}[th]
		\begin{center}
			\includegraphics[width=\linewidth]{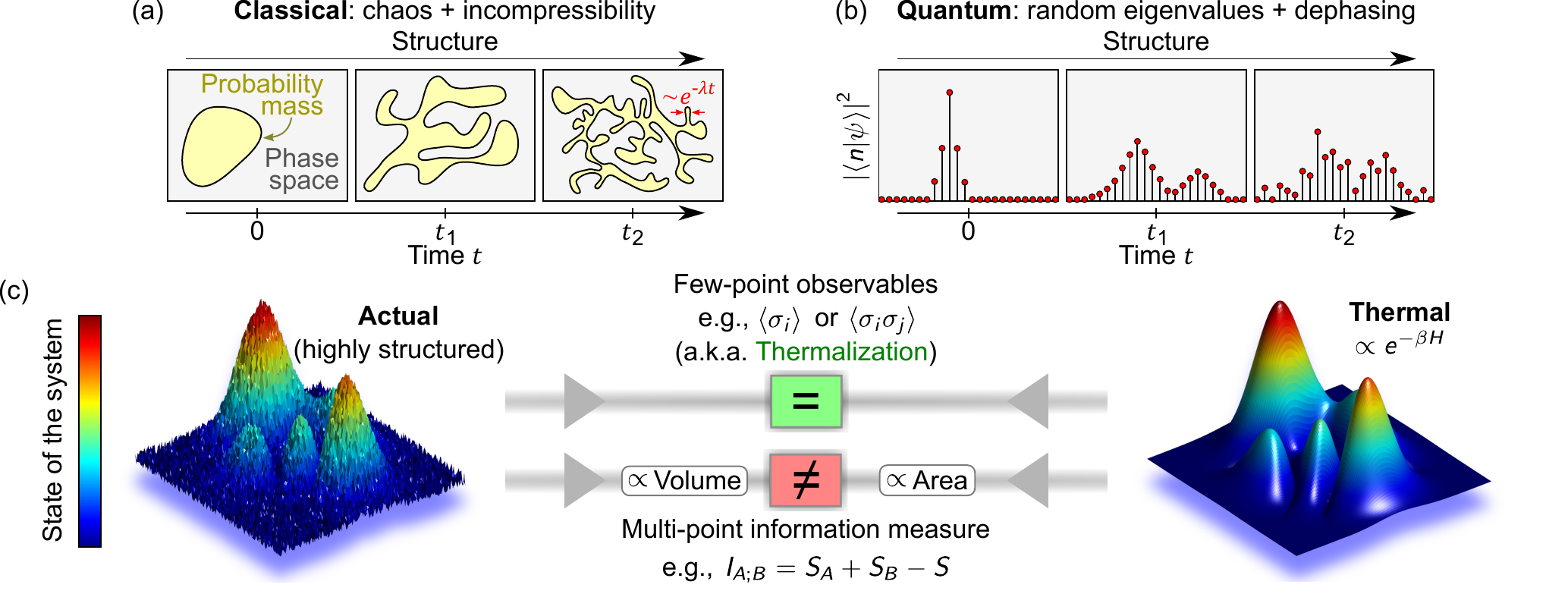}\\
		\end{center}
		\vskip -0.5cm \protect
		\caption{
			\textbf{Emergence of highly structured states in classical and quantum many-body dynamics.}
			(a) Schematic representation of the evolution of the probability mass in the classical phase space under Hamiltonian dynamics. Due to chaos and incompressibility, the probability mass becomes more and more structured in phase space, with fine features of size $\sim e^{-\lambda t}$, with $\lambda$ the Lyapunov exponent, effectively acquiring a random character.
			(b) Schematic representation of the evolution of the wavefunction, represented with respect to a computational basis $\{\ket{n}\}$ sorted with respect to $\langle n | H |n \rangle$. The wavefunction becomes more and more structured due to the dephasing effects from an effectively random spectrum.
			(c) The state of the system, whether in the form of a classical probability distribution or quantum wavefunction, is at long times very highly structured in generic many-body dynamics, as pictorially represented on the left. By contrast, a thermal distribution $\propto e^{-\beta H}$ washes out much of this fine structure (right). If the system has thermalized, few-body observables such as order parameters or correlation functions are correctly predicted by the thermal distribution $\propto e^{-\beta H}$ (top), but multi-point information measures such as the MI are not (bottom). Specifically, thermal states have area-law MI whereas the exact state at late times has volume law. It is thus necessary to keep track of the exact state of the system to correctly account for information spreading, which is clear in the quantum setting, and which we here consider in the classical one.}
		\label{cEEfig1}
	\end{figure*}
	
	\section{A lesson from quantum mechanics: the importance of tracking the global state}
	\label{sec. thermalization}
	
	Touching upon a number of well-known ideas around the concepts of thermalization and convergence of ensembles, in this Section we establish close parallels between classical and quantum quench dynamics. This highlights the need to account for the global state of the system when looking at the MI, a key conceptual step necessary to correctly frame the classical problem on par with the quantum one.
	
	Consider a many-body system undergoing non-dissipative dynamics generated by the Hamiltonian $H$. The global state of the system is described by a wavefunction $\ket{\psi}$ and by a probability distribution $p$ in the quantum and classical cases, respectively~\footnote{Strictly speaking, the classical system is described by a continuous probability density on phase space. This will be discussed later in Section~\ref{sec. Hamiltonian dynamics}, but should not disturb the reading now}. The wavefunction is represented with respect to a certain basis $\{\ket{n}\}$ as $\ket{\psi} = \sum_n \bra{n} \ket{\psi} \ket{n}$, whereas the probability distribution is represented with respect to a certain set of coordinates $x$ as $p(x)$. We are interested in the general scenario of local, many-body, and non-integrable Hamiltonians $H$.
	
	We assume the standard situation in which fluctuations are initially local, meaning that the various parts of the system are statistically independent. Specifically, we consider a product state $\ket{\psi(0)} = \ket{\psi_A(0)} \otimes \ket{\psi_B(0)}$ and a disjoint probability distribution $p(0) = p_{A}(0)p_{B}(0)$ for a quantum and a classical system, respectively, such that $I_{A;B}(0) = S_e(0) = 0$. The initial quantum wavefunction (classical probability distribution) has a simple aspect in the Hilbert (phase) space. Of course, the shape depends on the specific choice of the basis (coordinates), but can be pictured as something simple if the latter is local, which we shall assume, see left panels in Fig.~\ref{cEEfig1}(a,b).
	
	Under the non-dissipative dynamics, the spreading of correlations breaks the local character of the wavefunction (probability distribution), which in the local basis becomes therefore more and more `structured'. In the quantum case, this can be understood from the wavefunction amplitudes $\left| \braket{n}{\psi(t)} \right|^2 = \left| \sum_E \braket{n}{E} e^{-i\frac{Et}{\hbar}} \braket{E}{\psi(0)} \right|^2$ acquiring an effectively random character due to the chaotic spectrum $\{E\}$ of many-body non-integrable Hamiltonians~\cite{Deutsch1991, Srednicki1994, rigol2008thermalization, Deutsch2018, atas2013distribution}. Indeed, the phases $e^{-i\frac{Et}{\hbar}}$ at long times $t$ become effectively random, resulting in probabilities $\left| \braket{n}{\psi} \right|^2$ that fluctuate in time and with no clear dependence on $n$, see the rightmost panel of Fig.~\ref{cEEfig1}(b). In the classical case, the dynamical randomization of the probability is instead due to incompressibility and chaos, which taken together imply the probability distribution to develop finer and finer features~\cite{kardar2007statistical}, or more and more structure, as time increases -- see the leftmost panel of Fig.~\ref{cEEfig1}(a).
	
	It turns out that these fine features make only little difference for most observables of interest, according to the paradigm of thermalization. The latter stipulates that, in the time evolution after a quench, local few-point observables equilibrate at long times to their thermal value, which depends on the initial condition only through its energy. This general idea has long been established in classical physics in terms of chaos and ergodicity~\cite{kardar2007statistical} and more recently for closed quantum many-body systems in terms of the eigenstate thermalization hypothesis (ETH)~\cite{Deutsch1991, Srednicki1994, rigol2008thermalization, Deutsch2018}. The concept of thermalization is naturally associated with that of convergence of ensembles~\cite{kardar2007statistical}, according to which we can think of the distribution of the system (classical probability $p$ or quantum density matrix $\rho = \ket{\psi}\bra{\psi}$) as approaching the stationary thermal one $\propto e^{-\beta H}$.
	
	This point deserves much care, though. As a matter of fact, the idea of convergence of the ensembles only holds at the level of few-point observables, reduced density matrices, and marginal probability distributions. But as described above, the global state itself is far from stationary, let alone thermal, and this can have deep consequences on many-point observables. Most importantly, while the long-time value of the quantum MI fulfills volume law~\cite{calabrese2005evolution, kim2013ballistic}, which is rooted precisely in the effective randomness of the many-body wavefunction~\cite{page1993average}, the MI of a thermal state fulfils area-law scaling~\cite{wolf2008area}. Indeed, a thermal distribution $\rho_{th}$ washes out the fine structure of the state, thus missing a large (extensive) amount of MI, as pictorially illustrated in Fig.~\ref{cEEfig1}(c).
	
	This example highlights the importance of keeping track of the exponentially large state of the system for studying many-body information spreading. While this is customary for quantum theories, it is not for classical ones, that instead either assume its convergence to a canonical ensemble, or focus on few-body observables within Monte-Carlo sampling. Individuating this difference as the main origin of the gap between our understanding of classical and quantum information spreading is the main conceptual finding of our work.
	
	Before substantiating these ideas with numerics on Hamiltonian dynamics, which involves an extra phase-space discretization procedure and that we postpone to Section~\ref{sec. Hamiltonian dynamics}, let us take a step back and start by considering the simpler case of classical cellular automata.
	
	\begin{figure*}[th]
		\begin{center}
			\includegraphics[width=\linewidth]{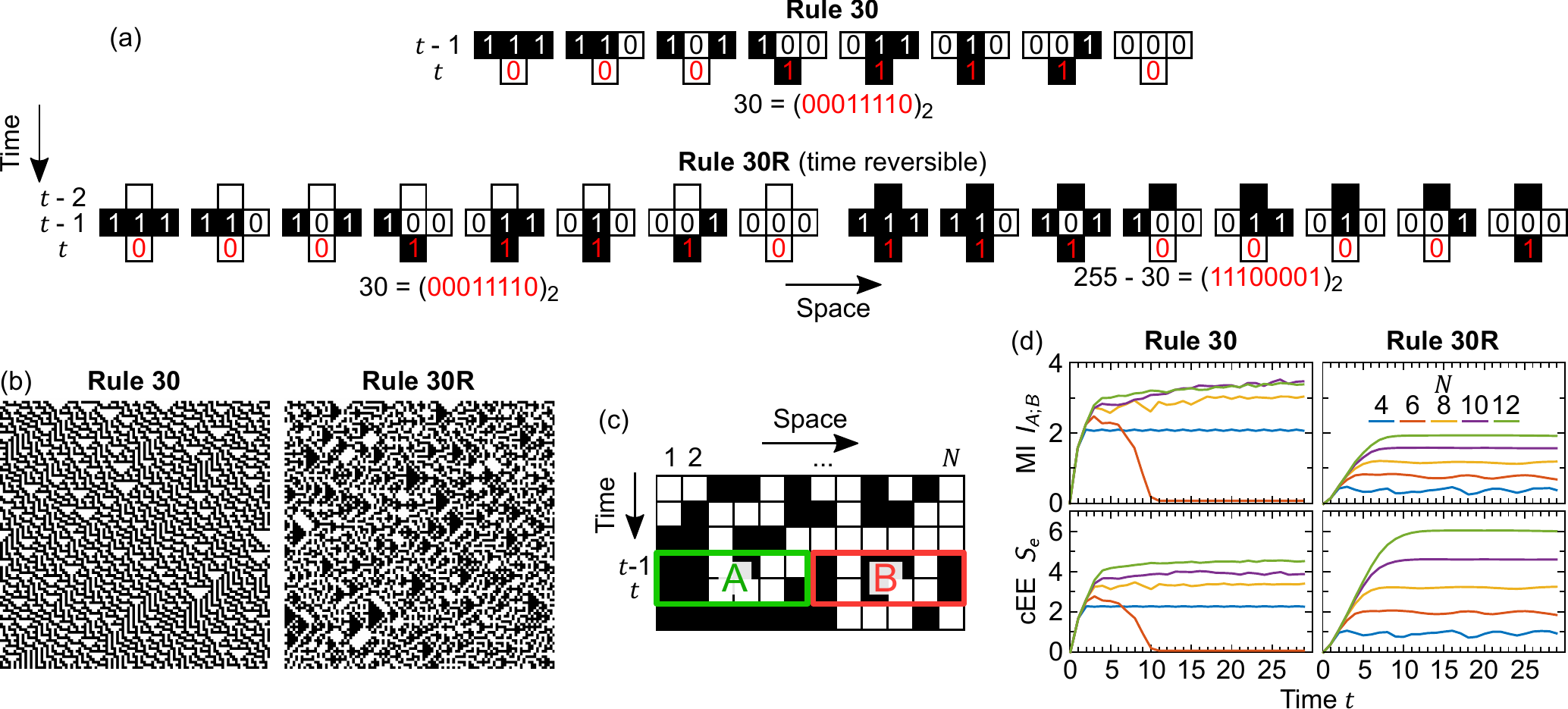}\\
		\end{center}
		\vskip -0.5cm \protect
		\caption{
			\textbf{Many-body information spreading in classical cellular automata.}
			(a) Rule 30 cellular automaton (top), and its (second-order) reversible version Rule 30R (bottom).
			(b) Instances of the space-time profiles of Rule 30 (left) and Rule 30R (right).
			(c) Graphical representation of the subsystems $A$ and $B$ that are used to compute MI and cEE. The state of the system at time $t$ is unambiguously defined by the state of the bits at time $t$ and at time $t-1$.
			(d) Dynamics of MI and cEE for various system sizes $N$. In Rule 30 (left), the dynamics of MI and cEE does not exhibit typical features of classical and quantum non-dissipative dynamics. By contrast, Rule 30R captures features such as the initial linear growth of entanglement $\sim t$ and its eventual saturation to an extensive $\sim N$ value (reached after a time $t \sim N$). Indeed, this stresses that key ingredients behind this phenomenology are locality, chaos, and time-reversibility, the latter of which is featured by Rule 30R but not by Rule 30. Here, we used $q_0 = 0.7$.}
		\label{cEEfig2}
	\end{figure*}
	
	\section{Information spreading in classical cellular automata}
	\label{sec. CA}
	
	In this Section, we apply our probabilistic framework to investigate information spreading in classical cellular automata. Imagine to draw an initial condition from a disjoint (in $A$ and $B$) initial probability distribution and to evolve it ``blindly'', meaning without looking at it. If at time $t$ we inspect the state of half of the system $A$, how much do we learn about $B$? To address this question, we will consider how the many-body probability distribution evolves under the automaton dynamics and compute the MI and cEE. In addition to offering a discrete setting convenient for implementation, cellular automata help us to highlight the role played by time reversibility in the dynamics of MI and cEE.
	
	Consider a system made of $N$ bits $\bm{s} = (s_1, s_2, \dots, s_N) \in \{0,1\}^N$. The bit-string evolves in time under the action of some update rule, $\bm{s}(t+1) = \text{Rule}\left[\bm{s}(t)\right]$, with discrete time $t = 0,1,2,\dots$. For concreteness, we shall henceforth focus on Wolfram's Rule 30~\cite{wolfram2002}. This rule shares two key features with the standard quantum settings for information spreading, namely chaos and locality, but lacks a third: time reversibility. This ingredient is decisive, because it ensures information to be preserved in time. In fact, of the 256 rules that can be obtained from local updates involving only the nearest neighbouring sites, none is both chaotic and time-reversible~\cite{wolfram2002}. One way of recovering time-reversibility from a given Rule, while preserving chaos, is to modify it as $\bm{s}(t+1) = \text{Rule}\left[\bm{s}(t)\right] \bigotimes \bm{s}(t-1)$, where $\bigotimes$ denotes a bit-wise logical XOR operation~\cite{wolfram2002,aldana2003boolean}. Such a modified rule, that we shall call Rule R, constitutes a `second-order' automaton, in which the state of the system at time $t+1$ does not just depend on the state of the system at time $t$, but also on that at time $t-1$. Put differently, the reversibility of the automaton means that there exist an injective map connecting $\bm{s}(1)$ and $\bm{s}(0)$ to $\bm{s}(t)$ and $\bm{s}(t-1)$, and vice versa. Rule 30, Rule 30R, and two instances of the spatio-temporal profiles that they generate are shown in Fig.~\ref{cEEfig2}(a,b).
	
	In the case of first-order automata like Rule 30, the full information on the system is carried, at all times $t$, by a probability distribution $p$ with support on the $2^N$ possible bit-string configurations. In the case of a second-order automaton like Rule 30R, however, a similar distribution does not suffice. Indeed, what matters in this case is not just the configuration of the bit-string at time $t$, but also that at time $t-1$. Consequently, to retain the full information on the system, one has to define the possible states $\bm{\sigma}$ as the $2N$ bit-strings composed of the bits $\bm{s}(t)$ and $\bm{s}(t-1)$, see Fig.~\ref{cEEfig2}c. The probability distribution $p$ will correspondingly have $2^{2N}$ components~\footnote{Note that to simplify numerics one could alternatively consider, also for second-order cellular automata like Rule 30R, a shorter probability distribution defined on the $2^N$ single-time $N$-bit strings only. This would correspond to a phase-space reduction, which is sensible provided that only $2^N$ possible initial conditions are considered (e.g., by setting $s_j(t=-1) = 0$ while allowing for random $s_j(t=0)$). This idea will become clearer in Section~\ref{sec. Hamiltonian dynamics} for Hamiltonian dynamics, where such a reduction is necessary for phase-space discretization purposes. While this is possible, we have adopted a cellular automaton here precisely so we do not to have to deal with any phase-space reduction.}.
	
	Initially, we take the bits $\{s_j(t=0)\}$ and $\{s_j(t = -1)\}$ to be independent and identically distributed (i.i.d.) random variables, equal to $1$ with probability $q_0$ and to $0$ otherwise. That is, each of the $4^N$ possible initial conditions $\bm{\sigma}_0$ has a probability $p_{\bm{\sigma}_0}(0)$ that reads
	\begin{equation}
	p_{\bm{\sigma}_0}(0) = \prod_{j = 1}^{2N}
	\left[ \sigma_{0,j} q_0 + (1-\sigma_{0,j}) (1-q_0) \right].
	\label{eq. CA q0}
	\end{equation}
	
	The evolution of the probability vector from one time to the next is described by a map,
	\begin{equation}
	p(t+1) = F\left[p(t)\right],
	\label{eq. F}
	\end{equation}
	which can be determined once and for all by checking how each two-time microstate $\bm{\sigma}$ is evolved by one application of the automaton. We note that, in the case of reversible automata, like Rule 30R, the map $F$ acts as a permutation of the elements of $p$, because of injectivity. Indeed, in this case the probability does not change when ``sitting'' on a discrete trajectory $\bm{\sigma}_t$ starting in $\bm{\sigma}_0$. In a formula, $p_{\sigma_t}(t) = p_{\sigma_0}(0)$, which we can view as the discrete version of the Liouville's theorem for Hamiltonian systems~\cite{kardar2007statistical}.
	
	By iterating Eq.~\eqref{eq. F}, we evolve the many-body probability distribution $p(t)$ and investigate the corresponding MI and cEE dynamics for various system sizes $N$ in Fig.~\ref{cEEfig2}(d). In the non-reversible case of Rule 30, we find that the MI and the cEE generally grow until reaching a stationary value (unless in the case $N = 6$, which appears somewhat special and that we attribute to particularly severe finite-size effects). The key features characterizing quantum information spreading, namely linear growth and extensivity at short and long times, respectively, are absent. In striking contrast, these instead appear in the time-reversible automaton. In this case, we find that both the MI and the cEE grow linearly in time, $S_e,I_{A;B} \propto t$ until a time $\propto N$, after which saturation to an extensive value $\propto N$ takes over. Since the cEE is the EE obtained by treating the classical probability as a wavefunction, see Eq.~\eqref{eq. classical wavefunction}, we understand that the extensivity of the asymptotic cEE is due to an effective randomization of $p$, analogously to how in the quantum case it emerges from the effective randomness of the wavefunction~\cite{page1993average}. Classical cellular automata therefore allow to appreciate time-reversibility as a key ingredient underpinning the distinctive features of classical and quantum information spreading.

	\section{Measurement-induced transition}
	\label{sec. measurements}
	
	A natural question thus emerges from the previous Section, as it did in the quantum case: how does the tendency of chaotic dynamics to build extensive MI and cEE compete with the disentangling effect of local measurements? We turn to this question now, and show a classical MIPT, in which the rate of local measurements determines whether the asymptotic MI and cEE fulfill volume- or area-law scaling, in close analogy to its quantum counterpart~\cite{skinner2019measurement}.
	
	\begin{figure*}[th]
		\begin{center}
			\includegraphics[width=\linewidth]{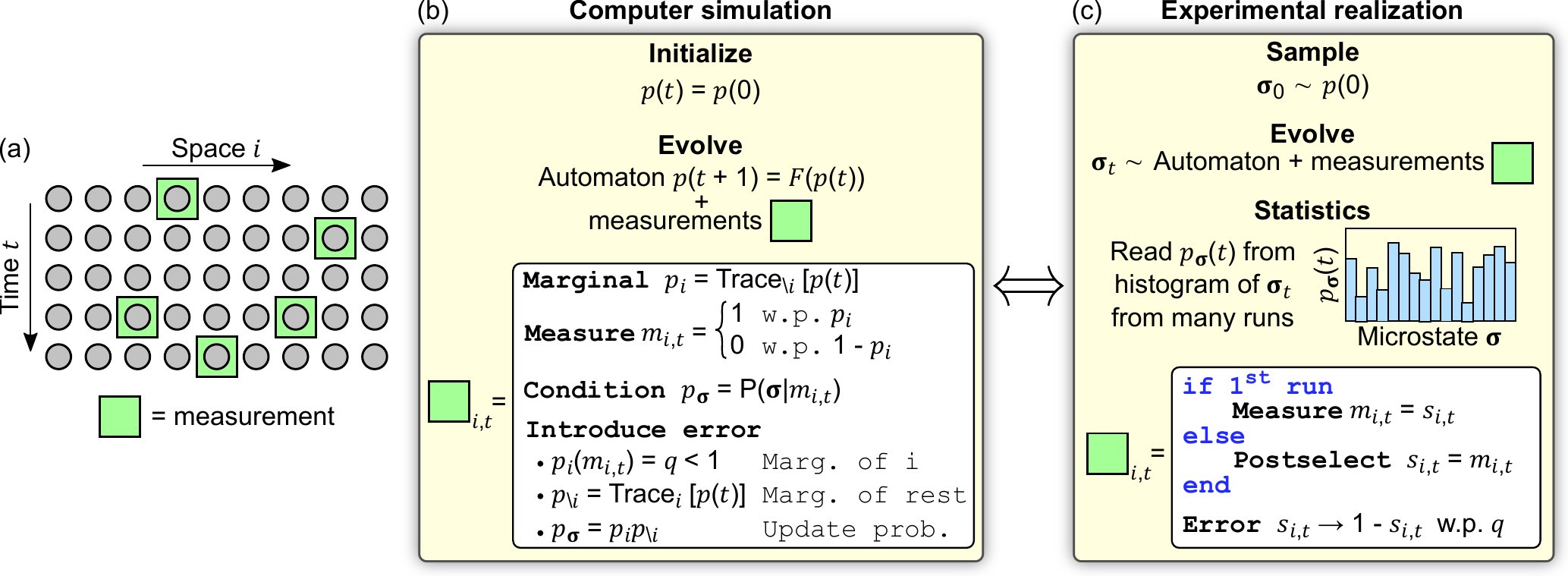}\\
		\end{center}
		\vskip -0.5cm \protect
		\caption{
			\textbf{Classical automaton with measurements -- A schematic protocol illustration.}
			(a) Each spacetime point can be measured (green boxes) with probability $p_m$.
			(b) The corresponding evolution of the many-body probability dynamics $p(t)$ can be obtained in a computer simulation. The probability is initialized as $p(t) = p(0)$, and evolved under the automaton dynamics according to the map $p(t+1) = F\left[p(t)\right]$ in Eq.~\eqref{eq. F}. Measuring site $i$ at time $t$ corresponds to (i) computing the marginal distribution of the site $p_i$, (ii) drawing a measurement outcome according to $p_i$, (iii) conditioning the probability accordingly, (iv) accounting for single-site fluctuations from a faulty measurement apparatus that flips the measured spin with probability $q$. The last point is in practice achieved by transforming the post-measurement marginal from $p_i(m_{i,t}) = 1 - p_i(1-m_{i,t}) = 1$ to $p_i(m_{i,t}) = q<1$, computing the marginal of the rest of the system $p_{\setminus i}$, and reconstructing the total distribution as $p = p_i p_{\setminus i}$. (c) Alternatively, the probability distribution could be obtained in an (numerical or actual) experiment by running the automaton on exponentially many initial configurations $\bm{\sigma}_0$ sampled from $p(0)$. The probability distribution $p(t)$ at later times is obtained as a histogram of the trajectories over the exponentially many possible microstates $\bm{\sigma}$. In the first run, measuring site $i$ at time $t$ consists of observing the state $s_{i,t}$ of the measured bit and storing it in the measurement variable $m_{i,t}$. For all subsequent runs, the measurements are rather postselections: if $s_{i,t}$ does not match the $m_{i,t}$ measured in the first run, the trajectory is discarded. In any run, the error possibly introduced by the faulty measurement apparatus is then accounted for by flipping the state of the measured bit with probability $1-q$. Remarkably, as argued in Section~\ref{sec. cl vs qu}, both the procedures in (b) and (c) are essentially identical to those needed in the quantum setting.}
		\label{cEEfig3}
	\end{figure*}
	
	\subsection{Protocol and classical measurements}
	
	Imagine as in Section~\ref{sec. CA} to sample an initial condition $\bm{\sigma}_0$ and evolve it with the automaton. As long as we do not look at it, the state of the automaton at time $t$ is described by a probability distribution $p(t)$, that starts from $p(0)$ as in Eq.~\eqref{eq. CA q0}, and evolves as $p(t+1) = F\left[p(t)\right]$ as in Eq.~\eqref{eq. F}. Information spreading generally builds MI between a given site $j$ and the rest of the system, meaning $p \neq p_j p_{\setminus j}$ and $I_{j; \setminus j} > 0$, with $p_j = \Tr_{\setminus j} p$ and $p_{\setminus j} = \Tr_j p$ the marginals of site $j$ and of the rest of the system $\setminus j$, respectively. Now, imagine that the `blind' evolution of the automaton is interleaved with random local measurements: each spacetime point $(j,t)$ can be observed with probability $p_m$, Fig.~\ref{cEEfig3}(a). Measuring the state of site $j$ at time $t$ reduces our ignorance on the system and conditions the probability distribution~\cite{cover2006elements}
	\begin{equation}
	p_{\bm{\sigma}}(t^+) = 
	\begin{cases}
	\frac{p_{\bm{\sigma}}(t^-)}{p_j(m_{j,t},t^-)} \ & \text{if} \ \sigma_j = m_{j,t} \\
	0 \ & \text{if} \ \sigma_j \neq m_{j,t}.
	\end{cases}
	\label{eq. Conditioning}
	\end{equation}
	In Eq.~\eqref{eq. Conditioning}, $t^-$ and $t^+$ refer to the instants before and after the measurement, respectively, $m_{j,t}$ is the measurement outcome, and $p_j(m_{j,t},t^-)$ its a priori probability. Simply put, to obtain $p(t^+)$, we remove all states incompatible with the measurement outcome and normalize the distribution again. In this process, the marginal of site $j$ becomes $p_{\sigma_j}(t^+) = \delta_{\sigma_j, m_{j,t}}$, implying that the post-measurement conditional many-body probability factorizes, $p(t^+) = p_j(t^+) p_{\setminus j}(t^+)$, and that the MI between site $j$ and the rest of the system vanishes, $I_{j;\setminus j} = 0$.
	
	After many measurements, the suppression of probability components in Eq.~\eqref{eq. Conditioning} would eventually lead at long times to the collapse of the state of the system to a single microstate $\bm{\sigma}^*$, $p_{\bm{\sigma}} \to \delta_{\bm{\sigma}, \bm{\sigma}^*}$, for which no information can spread at all. More interesting is the situation in which the measurement apparatus is faulty, and acts as a source of fluctuations. Specifically, we suppose that, after a measurement has been performed, the measured bit randomly flips with an error probability $1-q$, with $0<q<1$. We use $t^{++}$ to refer to the time right after the possible mistake has been introduced. We have that
	\begin{equation}
		p_{\sigma_j}(t^{++}) = q \delta_{\sigma_j, m_{j,t}} + (1-q) \delta_{\sigma_j, 1-m_{j,t}}.
		\label{eq. measurement error}
	\end{equation}
	The many-body probability distribution is modified accordingly, $p(t^{++}) = p_j(t^{++}) p_{\setminus j}(t^+)$.
	
	The local measurement protocol is closely analogous to the quantum case. There, the $j$-th qubit is sampled according to its reduced density matrix $\rho^{(j)} = \Tr_{\setminus j} \rho$: the qubit takes value $1$ with probability $\rho_{1,1}^{(j)}$, and value $0$ with probability $\rho_{0,0}^{(j)}$. The post-measurement fluctuations at site $j$ are intrinsic and due to Heisenberg uncertainty. As in the classical case, the effect of a quantum measurement is that of `factoring out' the state of $j$ from that of the rest of the system, meaning $\ket{\psi(t^+)} = \ket{\psi_j(t^+)} \ket{\psi_{\setminus j}(t^+)}$ or equivalently $\rho(t^+) = \rho_j(t^+) \rho_{\setminus j}(t^+)$, all of which are conditional on the measurement outcome.
	
	\begin{figure*}[th]
		\begin{center}
			\includegraphics[width=\linewidth]{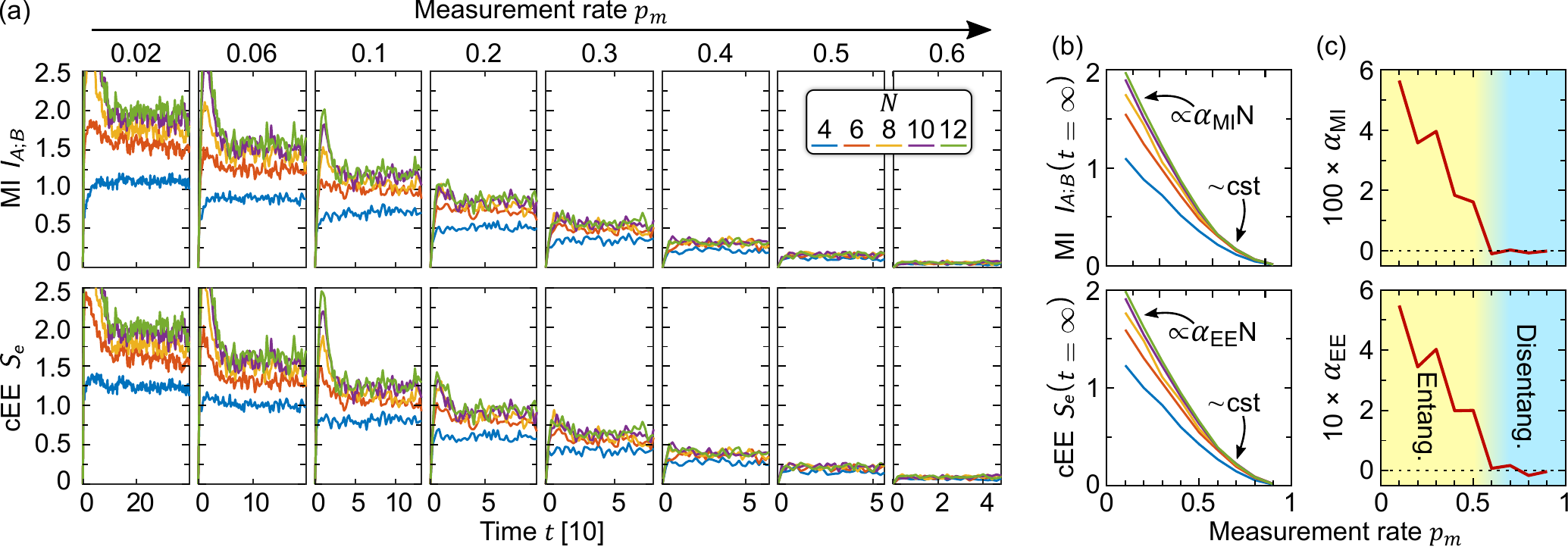}\\
		\end{center}
		\vskip -0.5cm \protect
		\caption{
			\textbf{Classical measurement-induced phase transition.}
			(a) MI and cEE dynamics for various measurement rates $p_m$ and system sizes $N$. After a very quick transient (partially cut), the MI and cEE saturate to a stationary value, up to temporal fluctuations due to finite-size effects. Crucially, the stationary value increases with $N$ if $p_m$ is small enough, and is insensitive of $N$ otherwise.
			(b) The long-time asymptotic values $I_{A;B}(t=\infty)$ and $S_e(t=\infty)$ are obtained by averaging the results in (a) over the second half of the shown time interval, and plotted versus $p_m$ for various system sizes $N$. The MI and cEE decay with $p_m$, reaching $0$ at $p_m = 1$.
			(c) The scaling coefficients $\alpha$, such that $I_{A;B}(t=\infty) \approx \alpha_{MI} N$ and $S_e(\infty) \approx \alpha_{EE} N$, are obtained through linear fits for and highlight an entangling-disentangling MIPT (the fitting excludes $N = 4$ and $6$ to reduce finite-size effects). For $p_m \lessapprox 0.5$ the scaling coefficients are finite, indicating extensivity of the asymptotic MI and cEE and thus unbounded growth in the thermodynamic limit. By contrast, for $p_m \gtrapprox 0.5$ the long-time values of MI and cEE become insensitive to system size $N$. Here, we used $q_0 = 0.8$, $q = 0.75$, and $R = 100$.}
		\label{cEEfig4}
	\end{figure*}

	The resulting evolution of the probability distribution can be determined in a computer simulation according to the procedure schematized in Fig.~\ref{cEEfig3}(b). To measure the probability distribution $p$ from an experiment, a procedure as in Fig.~\ref{cEEfig3}(c) is instead required, which unfolds as follows. An ensemble of initial states $\bm{\sigma}_0$ is sampled from the distribution $p(0)$, and evolved under the automaton rules. For the first of these runs, measuring a site $i$ at time $t$ means taking note of the state of $s_i(t)$ of the bit, and saving it as $m_{i,t}$. The measurement outcome $m_{i,t}$ is implicitly distributed according to the marginal $p_i$. The faulty measurement apparatus then introduces fluctuations as in Eq.~\eqref{eq. measurement error}, by leaving the bit $s_i(t)$ unchanged or inverted with probabilities $q$ and $1-q$, respectively. For all the subsequent runs, measurements always happen in the same spacetime locations, and are not measurements but post-selections. If $s_{i,t}$ does not match the $m_{i,t}$ measured in the first run, the trajectory is discarded, whereas otherwise the trajectory is maintained. Remarkably, this procedure is in close analogy to the one that should be followed in a quantum experiment, as we will argue in Section~\ref{sec. cl vs qu}. Repeating this procedure exponentially many times, the probability $p_{\bm{\sigma}}(t)$ is estimated as the fraction of trajectories found in $\bm{\sigma}$ at time $t$ and, in the limit of infinitely many runs, matches the probability obtained from the procedure in Fig.~\ref{cEEfig3}(b) discussed earlier.
	
	\subsection{Numerical results}
	
	The above protocol is implemented in Fig.~\ref{cEEfig4}. In Fig.~\ref{cEEfig4}(a) we plot the dynamics of MI and cEE for various measurement rates $p_m$. After an initial transient, whose duration is larger for smaller $p_m$ and that is partially cut by the chosen axes limits, the MI and the cEE reach a stationary value, up to finite-size temporal fluctuations. Asymptotic values for the MI and cEE are obtained as time averages and plotted in Fig.~\ref{cEEfig4}(b) versus the measurement rate $p_m$, and for various system sizes $N$. Strikingly, for small $p_m$, the asymptotic values of MI and cEE scale with the system size $N$, whereas they do not for large $p_m$. To make the last statement quantitative, we use a linear fit to extract the scaling coefficients $\alpha_{MI}$ and $\alpha_{EE}$, with $I_{A;B}(\infty) \approx \alpha_{MI} N$ and $S_e(\infty) \approx \alpha_{EE} N$, and plot them versus $p_m$ in Fig.~\ref{cEEfig4}(c). Indeed, these exhibit a transition from finite value to zero at $p_m \approx 0.6$, thus showing the first classical MIPT between area- and volume-law cEE and MI. The cEE closely follows the qualitative behaviour of the quantum EE, see e.g.~Fig.~13 in Ref.~\cite{skinner2019measurement}. In our simulations we take $q = 0.75$ and $q_0 = 0.8$ but the key qualitative features are not contingent on this choice. Generally, we do not expect the asymptotic MI and cEE (thus, the MIPT) to depend on the initial distribution.
	
    In this Section, we considered the most natural classical measurement procedure, that conditions the many-body probability distribution depending on the measurement outcome. We stress that similar results can be obtained for other local protocols that have the ability to decouple local fluctuations, that is, to suppress the mutual information between a local degree of freedom and the rest of the system, which can be achieved in many ways in both the classical and quantum settings. For instance, one could consider simple resettings, in which the measured bit is set to $1$ with probability $q$ and to $0$ with probability $1-q$. Away from the probabilistic framework adopted here, we shall also note that the idea of MIPT can be generalized to even simpler measurement protocols, e.g., defined as locally projecting two trajectories on one another~\cite{willsher2022measurement}.
	
	\section{Information spreading in many-body Hamiltonian dynamics}
	\label{sec. Hamiltonian dynamics}
	
	Thanks to their discrete nature, cellular automata have proven particularly convenient for numerical implementation. But let us now go back to the case of Hamiltonian dynamics, which we have already considered in Section~\ref{sec. thermalization} as a motivation for keeping track of the full many-body probability. Going beyond the result for area-law MI in a thermal ensemble~\cite{wolf2008area}, in this Section we find a simple expression for the asymptotic MI in Hamiltonian dynamics. We find it to be generally volume-law, reducing to area-law only at an infinite effective temperature, unlike the cEE. Introducing a careful phase-space discretization procedure we then numerically investigate information spreading in an interacting classical spin chain. We recover the salient features of information spreading in quantum quenches, and highlight the key differences between MI and cEE.
	
	\subsection{Asymptotic mutual information in Hamiltonian dynamics}
	\label{subsec. MI in Hamiltonian dynamics}
	First, we provide a heuristic expression for the asymptotic long-time MI in many-body Hamiltonian dynamics. First of all, Hamiltonian dynamics involves continuous variables, and thus a continuous probability density $p$ over the phase space. The first point to clarify is therefore how to make sense of the definitions in Section~\ref{sec. theory}, that are for discrete distributions instead. A discretization of the phase space can be performed dividing it into little volumes $\Delta$. In the continuous limit $\Delta \to 0$, it is straightforward to show that $S \approx \mathcal{S} - \log \Delta$, with $\mathcal{S} = \int dx \ p(x) \log(p(x))$ the differential entropy, for which we use a calligraphic notation and involving integration over the whole many-body phase space~\cite{cover2006elements}. Contrary to what one might naively expect, the differential entropy $\mathcal{S}$ is therefore \emph{not} the continuous limit of the discrete entropy $S$, which diverges as $\log \Delta$ instead. Fortunately, this divergence does not affect the MI, which instead does have a well-defined continuous limit~\cite{cover2006elements}. Indeed, in the expression of the MI in terms of the entropies, Eq.~\eqref{eq. MI}, the diverging terms of the entropies cancel, meaning that one can equivalently write $I_{A;B} = \mathcal{S}_{A} + \mathcal{S}_{B} - \mathcal{S}$ or $I_{A;B} = S_{A} + S_{B} - S$, in the limit $\Delta \to 0$.
	
	This remark being made, we can now find an expression for the long-time asymptotic value of the MI in Hamiltonian dynamics. Due to Liouville's theorem, the integral of any function of $p$ over the whole phase space is constant, $\frac{d}{dt}\int dx \ f(p(x)) = 0$. In particular, the differential entropy is conserved, $\mathcal{S}(t) = \mathcal{S}(0)$. While, strictly speaking, the total probability distribution does not thermalise due to chaos and incompressibility, see Section~\ref{sec. thermalization}, we expect the marginals to do so \footnote{The continuous marginal $p_A(A) = \int dB \ p(A,B)$ can be split as a sum of integrals over small volumes $\{\mathcal{B}_n\}$ of phase space at constant $A$, $p_A(A) = \sum_n \int_{\mathcal{B}_n} dB \ p(A,B)$. One can increase the number of volumes $\mathcal{B}_n$ to make their volume small enough, so that the energy $H$ within each of them can be considered constant (that is, with variation small compared to the involved energy scales). The probability distribution $p(A,B)$ fluctuates over a phase space length scale $\sim e^{-\lambda t}$ that becomes arbitrarily small and eventually much smaller than that of $\mathcal{B}_n$. Having arbitrarily many fluctuations within $\mathcal{B}_n$, we expect the probability to self-average, meaning that $\int_{\mathcal{B}_n} dB \ p(A,B) \approx \int_{\mathcal{B}_n} dB \ p_{th}(A,B)$. Rebuilding the integral one thus gets $p_A(A) = \int dB \ p_{th}(A,B)$, that is, that the marginal thermalizes.}. As a result, we can assume at sufficiently long times $t$ that $S_A(t) + S_B(t) \approx S^\beta + s_{al}$, where $S^\beta$ denotes the entropy of a thermal ensemble, whereas $s_{al}$ is an area-law correction. The inverse temperature $\beta$ is found from the usual consistency condition $\langle H \rangle_0 = \langle H \rangle_\beta$, that is, $\int dx \ p(t = 0,x) H(x) = \int dx \ z^{-1}e^{-\beta H(x)} H(x)$, with $z = \int dx  e^{-\beta H(x)}$ the partition function. From Eq.~\eqref{eq. MI} we thus obtain that the MI at sufficiently long times reaches the asymptotic value
	\begin{equation}
	I_{A;B}(\infty) \approx \mathcal{S}^\beta - \mathcal{S}(0) + s_{al}.
	\label{eq. asymptotic MI}
	\end{equation}
	
	Eq.~\eqref{eq. asymptotic MI} states that the asymptotic MI between two halves $A$ and $B$ of a many-body Hamiltonian system is equal to the difference between the initial entropy and a thermal entropy at temperature $\beta^{-1}$ set by the initial state $p(0)$, up to a boundary correction $s_{al}$. Note that the entropies in Eq.~\eqref{eq. asymptotic MI} can equivalently be meant as differential entropies or as discrete entropies in the continuous limit, since the diverging terms cancel anyway, as discussed above.
	
	An important observation is that, by definition, the thermal distribution is the one that maximises the entropy at a given temperature, and so $\mathcal{S}^\beta \ge \mathcal{S}(0)$, which is consistent with the positivity of the MI and in which the equality holds only for initial states with infinite temperature, for which $\beta = 0$ and $\langle H(x) \rangle_0 = \frac{\int dx \ H(x)}{\int dx }$, or for thermal initial distributions. Here, we use the term `infinite temperature state' in reference not to the infinite temperature thermal distribution, but to any probability distribution whose expected energy coincides with the one that the infinite temperature thermal distribution would predict. From Eq.~\eqref{eq. asymptotic MI} we can thus conclude that initial non-thermal distributions $p(0)$ corresponding to a finite effective temperature $0 < \beta < \infty$ generally lead to a volume-law asymptotic MI, $I_{A;B}(\infty) \approx \mathcal{S}^\beta - \mathcal{S}(0) = \mathcal{O}(N)$, whereas initial distributions $p(0)$ that are either thermal or at an effective infinite temperature $\beta = 0$ result in an area-law asymptotic MI, $I_{A;B}(\infty) \approx s_{al} \ll N$.
	
	\begin{figure*}[th]
		\begin{center}
			\includegraphics[width=\linewidth]{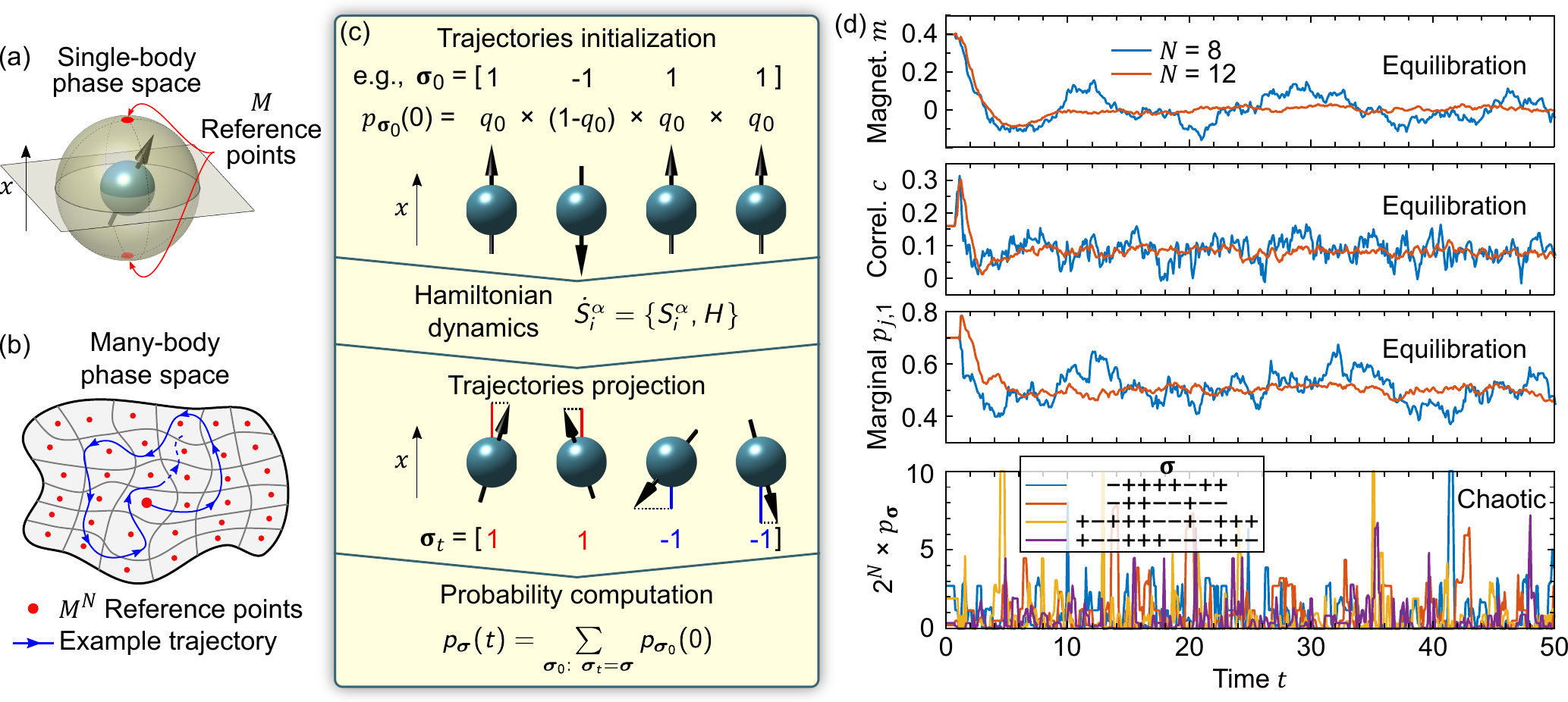}\\
		\end{center}
		\vskip -0.5cm \protect
		\caption{
			\textbf{Phase-space discretization for Hamiltonian dynamics.}
			(a) Discretization of the single-body phase space of a classical spin. The sphere is divided into two hemispheres, each identified by a reference point (at the pole, in red).
			(b) Discretization of the many-body phase space, that is divided into $M^N$ pockets, each of which is identified by a many-body reference point (red dots). A schematic trajectory starting from a reference point (larger red dot) is shown in blue.
			(c) Protocol for obtaining the dynamics of the discrete probability, in the illustrative case of $N = 4$ spins. As initial conditions, we consider the $2^N$ possible spin states in which each spin is either aligned or antialigned along $\bm{x}$ (corresponding to the reference points in (b)). Initial states are labelled with a bit-string $\bm{\sigma}$ and their associated probability $p_{\bm{\sigma_0}}$ factorizes. The trajectories are evolved under Hamiltonian dynamics, and their coordinates in the reduced phase space, obtained as $\bm{\sigma}_t = \text{sign}\left(S_1^x, \dots, S_N^x\right)$, are used to compute the probability distribution $p_{\bm{\sigma}}(t)$.
			(d) The ensemble averages of few-point observables, such as magnetization $m$, correlation $c$, and one-site marginal distribution $p_{\sigma_1}(t)$, equilibrate in time to the thermal value, up to temporal fluctuations that decrease with the system size $N$. By contrast, the many-body probability $p_{\bm{\sigma}}(t)$ itself does not equilibrate, independent of $N$. Here, we considered the spin model in Eq.~\eqref{eq. Heisenberg chain}, with parameters $J = 1$, $W = 2$, $R = 1$, and $q_0 = 0.7$.}
		\label{cEEfig5}
	\end{figure*}
	
	Further, to estimate the latter, we can compute the MI associated to a random probability distribution, in the same spirit of the calculation of the EE for a random wavefunction~\cite{page1993average}. Let us assume that the components of the probability $p_{\bm{\sigma}}$ are independent identically distributed (i.i.d.) random numbers with average $\langle p_{\bm{\sigma}} \rangle = \left[ \sum_{\bm{\sigma}} 1 \right]^{-1}$~\footnote{Strictly speaking, the components of the probability are not independent, because they are tied by the normalization $\sum_{\bm{\sigma}} p_{\bm{\sigma}}$. However, for $N \to \infty$, that is for a diverging number of possible microstates, the normalization constraint becomes weaker and weaker, and the components of the probability distribution can be considered as effectively i.i.d..}. For large $N$ and by virtue of the central limit theorem, the total entropy reads $S \approx - \frac{\langle p_{\bm{\sigma}} \log p_{\bm{\sigma}} \rangle}{\langle p_{\bm{\sigma}} \rangle}$. The marginals become instead uniform, $p_{\bm{\sigma}_A} \to \langle p_{\bm{\sigma}_A} \rangle$ and $p_{\bm{\sigma}_B} \to \langle p_{\bm{\sigma}_B} \rangle$, for which the marginal entropies are maximal, $S_A+S_B \approx - \log \langle p_{\bm{\sigma}} \rangle$. Plugging these results in Eq.~\eqref{eq. MI} quickly leads to the MI at infinite time and effective temperature
	\begin{equation}
	I_{A;B}^{\infty} \approx \left \langle \frac{p_{\bm{\sigma}}}{\langle p_{\bm{\sigma}} \rangle} \log \frac{p_{\bm{\sigma}}}{\langle p_{\bm{\sigma}} \rangle} \right \rangle.
	\label{eq. asymptotic and T = inf MI}
	\end{equation}
	
	Eq.~\eqref{eq. asymptotic MI}, Eq.~\eqref{eq. asymptotic and T = inf MI}, and the prediction of volume and area-law scaling of the asymptotic MI in many-body Hamiltonian systems at finite and infinite effective temperature initial conditions, respectively, are among the major findings of this work, which we now verify numerically.
	
	\begin{figure}[th]
		\begin{center}
			\includegraphics[width=\linewidth]{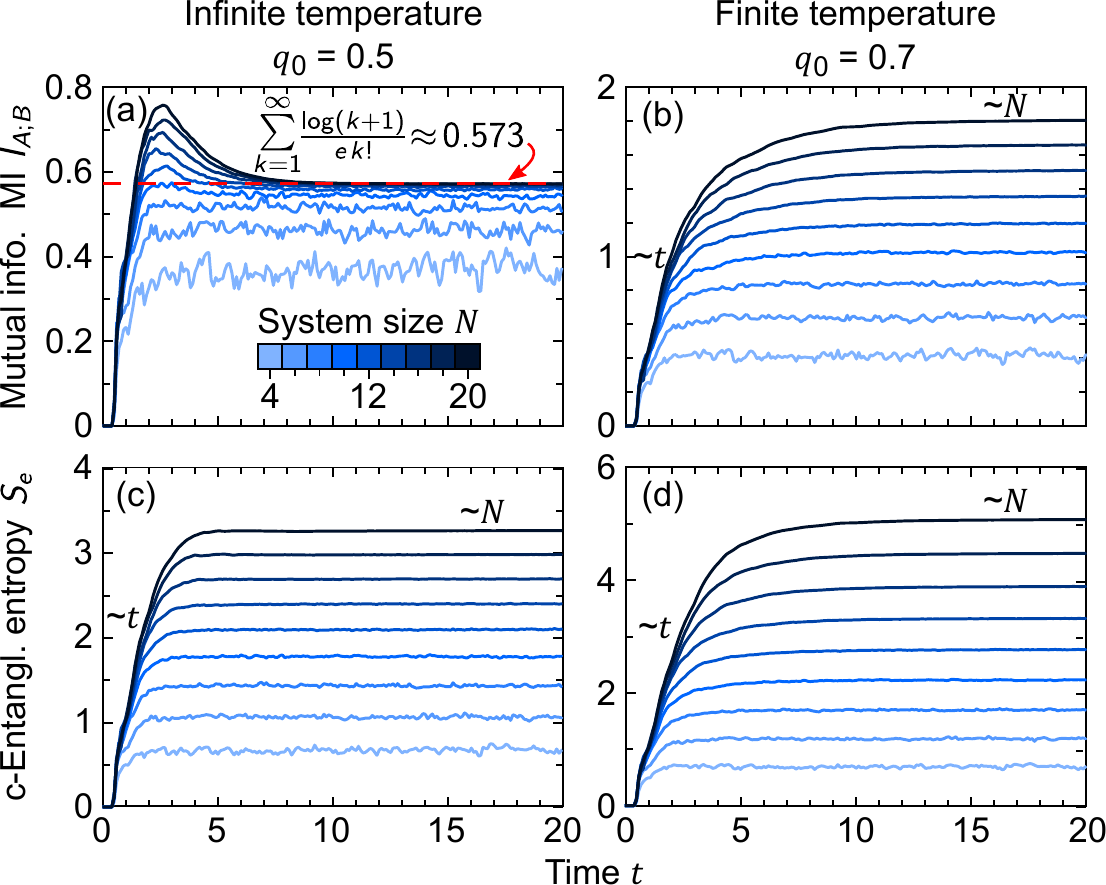}\\
		\end{center}
		\vskip -0.5cm \protect
		\caption{\textbf{Information dynamics in Hamiltonian many-body system.}
			The MI (top) and the cEE (bottom) are plotted versus time for various system sizes $N$. Two representative single-spin initial probabilities are considered: $q_0 = 0.5$ (left) and $q_0 = 0.7$ (right), corresponding to infinite and finite effective temperatures, respectively. In all the considered cases, the quantities of interest initially grow linearly in time signalling the ballistic spreading of correlations. At sufficiently long times, saturation to an asymptotic value sets in.
			(a) At infinite temperature, the asymptotic MI becomes insensitive to $N$ for $N \gtrapprox 10$, reaching the value predicted in Eq.~\eqref{eq. asymptotic MI Heisenb} and fulfilling area-law scaling, while (b) at finite temperature it fulfills a volume-law scaling instead. (c,d) On the other hand, the asymptotic cEE is extensive both at finite and infinite temperature.
			Here, we considered the Heisenberg chain in Eq.~\eqref{eq. Heisenberg chain} for $J = 1$, $W = 3$, and $R = 50$.}
		\label{cEEfig6}
	\end{figure}
	
	\begin{figure*}[th]
		\begin{center}
			\includegraphics[width=\linewidth]{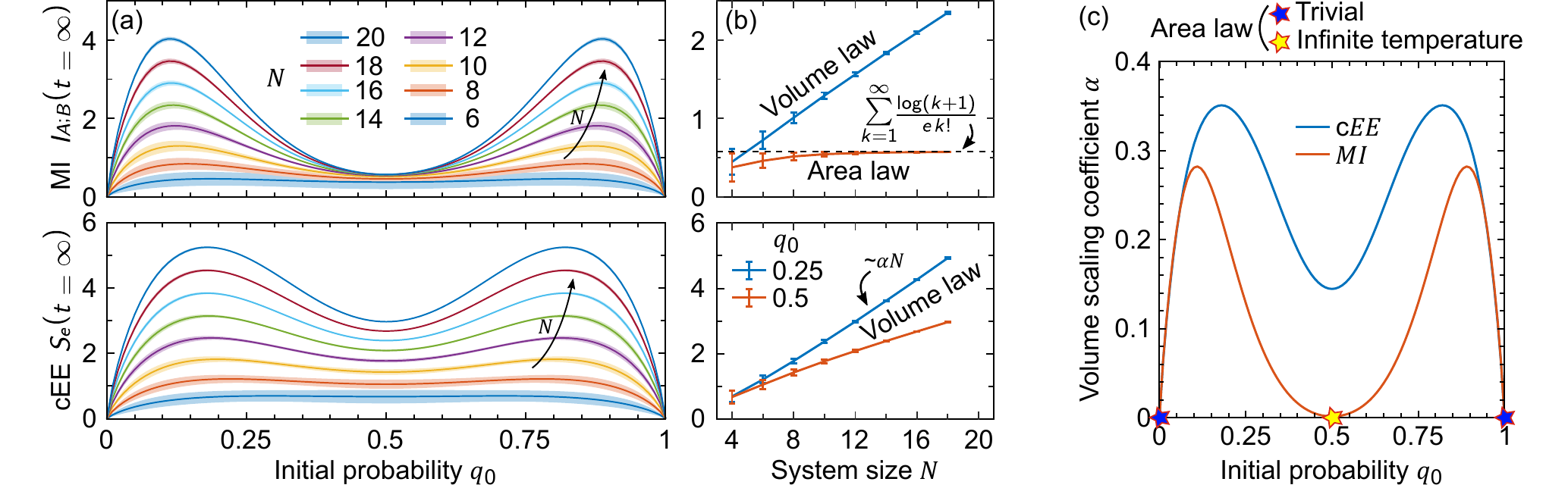}\\
		\end{center}
		\vskip -0.5cm \protect
		\caption{
			\textbf{Long-time scaling analysis in many-body Hamiltonian dynamics.} (a) The asymptotic MI $I_{A;B}(t = \infty)$ and cEE $S_e(t = \infty)$ are plotted versus the single-spin initial probability $q_0$ for various system sizes $N$. Solid lines and shaded bands represent mean and uncertainty (over the late time range $20<t<30$ and disorder realizations). At $q_0 = 0,1$ the dynamics freezes and no information spreads, leaving $I_{A;B} = S_e = 0$. At infinite temperature, $q_0 = 0.5,$ the asymptotic MI and cEE present local minima, but with a key difference: the former becomes independent of $N$ for $N \gtrapprox 10$, whereas the latter grows with $N$. For other values of $q_0$, both the asymptotic MI and cEE scale with $N$. (b) The scaling is better characterized for representative values of $q_0$, and found to be linear at large $N$, $\sim \alpha N$. The infinite-temperature value of the MI predicted in Eq.~\eqref{eq. asymptotic MI Heisenb} is reported for comparison. (c) The ultimate appreciation of the scaling properties is achieved by plotting the scaling coefficient $\alpha$ versus the initial probability $q_0$, as obtained from a linear fit (of the three largest values of $N$ only, to reduce small-size effects). Finite and vanishing values of $\alpha$ denote volume- and area-law scaling, respectively. The latter occurs at the trivial points $q_0 = 0,1$ and, only for the MI, at infinite temperature $q_0 = 0.5$. Here, we used $J = 1$, $W = 3$, and $R = 50$.}
		\label{cEEfig7}
	\end{figure*}
	
	\subsection{Case study: Heisenberg chain in a magnetic field}
	We numerically exemplify the above ideas for a paradigmatic model in the context of chaos, information spreading, and transport: the Heisenberg chain in a magnetic field. Consider $N$ classical spins $\bm{S}_j = \left(S_j^x, S_j^y, S_j^z\right)$ with unitary modulus $\left|\bm{S}_j\right|^2 = 1$ and $j = 1,2,\dots,N$, and with Hamiltonian
	\begin{equation}
	H = \sum_{j = 1}^N \left(\bm{S}_j \cdot \bm{S}_{j+1} + h_j S_j^z\right),
	\label{eq. Heisenberg chain}
	\end{equation}
	containing an isotropic nearest-neighbor interaction and a disordered magnetic field along $z$, the coefficients $\{h_j\}$ being independent random numbers drawn uniformly in $[-W,W]$. The system undergoes Hamiltonian dynamics, $\dot{S}_i^{\alpha} = \left\{S_i^\alpha, H\right\}$, where $\{\dots\}$ denotes Poisson brackets and $\left\{ S_i^\alpha, S_j^\beta \right\} = \delta_{i,j} \epsilon_{\alpha, \beta, \gamma} S_i^\gamma$, with $\delta_{i,j}$ the Kronecker delta, $\epsilon_{\alpha, \beta, \gamma}$ the Levi-Civita anti-symmetric symbol, and $\alpha,\beta,$ and $\gamma$ in $\{x,y,z\}$. Periodic boundary conditions are assumed, and results can be averaged over $R$ independent disorder realizations, as relevant. Again, the bipartition $(A,B)$ is chosen to correspond to the left and right halves of the chain ($N$ is assumed even with no loss of generality).
	
	The phase space of the system is that parametrized by $2N$ angles, two per spin, and is continuous as in any Hamiltonian system. As such, it requires discretization, which we perform in the minimal nontrivial way, that is, by splitting each solid angle in $M = 2$ hemispheres as illustrated Fig.~\ref{cEEfig5}(a). Two points that can be taken as representative for these hemispheres are the poles, which we may call `single-body reference points' and which we tag with a discrete variable $\sigma_j$ taking $M$ values, say $\sigma_j = \pm 1$ for the $\pm\bm{x}$ hemispheres, respectively. The many-body phase space is correspondingly divided into $M^N$ `pockets', each of which is represented by a many-body reference point (in which the spins points either towards $+\bm{x}$ or towards $-\bm{x}$), and tagged by a bitstring $\bm{\sigma} = \left(\sigma_1,\sigma_2,\dots,\sigma_N\right)$, see Fig.~\ref{cEEfig5}(b). Explicitly, the $\bm{\sigma}$-th pocket contains all the spin configurations such that $\text{sign} \left(S_1^x, S_2^x, \dots, S_N^x\right) = \bm{\sigma}$.
	
	Our goal is to find the dynamics of the discrete probability distribution $p_{\bm{\sigma}}(t)$. The idea is the following. (i) First, we consider the $M^N$ many-body reference points as possible initial conditions. Specifically, we consider each spin to initially point either along $+\bm{x}$ or along $-\bm{x}$, with probability $q_0$ and $1-q_0$, respectively, so that an initial configuration tagged $\bm{\sigma}_0$ has probability
	\begin{equation}
	p_{\bm{\sigma}_0} = \prod_{j = 1}^N
	\left[ \sigma_{0,j} q_0 + (1-\sigma_{0,j}) (1-q_0) \right].
	\label{eq. Hamiltonian q0}
	\end{equation}
	Again, the initial factorizability of the probability distribution makes our stochastic initial condition the one-to-one correspondent of quantum product states, and implies $I_{A;B}(0) = S_e(0) = 0$.
	(ii) Second, we evolve the $M^N$ trajectories integrating the ordinary differential equations (ODE) of Hamiltonian dynamics with an ODE solver, which necessarily makes them depart from the reference points they started from, and explore the continuous phase space. (iii) Third, each trajectory is projected onto the discrete space to define $\bm{\sigma}_t = \text{sign} \left[S_1^x(t), S_2^x(t), \dots, S_N^x(t)\right]$ at time $t$. (iv) Fourth, the probability $p_{\bm{\sigma}}(t)$ is obtained as
	\begin{equation}
	p_{\bm{\sigma}}(t)
	= \sum_{\bm{\sigma}_0: \bm{\sigma}_t = \bm{\sigma}} p_{\bm{\sigma}_0}(0),
	\label{eq. projection CA}
	\end{equation}
	that is, as the sum of the initial probabilities $p_{\bm{\sigma}_0}(0)$ of those states $\bm{\sigma}_0$ that, at a time $t$, are found in the $\bm{\sigma}$-th pocket of the phase space. (v) Fifth and final, the thus obtained vector of discretized probabilities $p$ is used to compute quantities of interest, including the MI and cEE.
	
	For the trivial points $q_0 = 0,1$ the spins are all initially perfectly polarized along either $-\bm{x}$ or $+\bm{x}$, respectively, and remain so at all times, with no dynamics happening, no spreading of correlations, $p(t) = p(0)$, and $I_{A;B} = S_e = 0$. On the other hand, the value $q_0 = 0.5$ corresponds to an infinite temperature state, $\langle H \rangle_0 = 0 = \langle H\rangle_{\beta = 0}$.
	
	First and foremost, we aim at verifying that the discretization protocol that we proposed leads to the desired key features of many-body dynamics, and in particular to the contrast between the equilibration of few-point observables and the non-equilibration of the probability itself, à la Liouville. In Fig.~\ref{cEEfig5}(c) we show, for increasing system sizes $N$, the dynamics of the magnetization $m(t) = \frac{1}{N} \sum_{\bm{\sigma}} p_{\bm{\sigma}}(t) \sum_{j = 1}^N \sigma_j$, of a correlator $c(t) = \frac{1}{N} \sum_{\bm{\sigma}} p_{\bm{\sigma}}(t) \sum_{j = 1}^N \sigma_j \sigma_{j+1}$, of a marginal probability $p_{\sigma_1 = 1}(t)$, and of the whole many-body probability $p_{\bm{\sigma}}(t)$ for some randomly chosen discretized states $\bm{\sigma}$. Few-point observables and marginal probabilities quickly equilibrate, with asymptotic temporal fluctuations as a finite-size result and decreasing with system size $N$ (even at a fixed and minimal discretization resolution $M=2$), whereas the many-body probability itself maintains (up to a factor $2^N$, due to normalization) chaotic fluctuations in time, without clear dependence on the state index $\bm{\sigma}$.
	
	These key features being verified, we are now in the position to finally investigate information spreading in many-body Hamiltonian dynamics through the lens of MI and cEE, whose time traces we plot in Fig.~\ref{cEEfig6} for various system sizes $N$ and initial single-spin probability $q_0$. Initially vanishing, both the MI and cEE grow linearly in time signalling the ballistic spreading of correlations, before saturating to an asymptotic value.
	
	Let us first focus on the MI. At infinite temperature ($q_0 = 0.5$), we find that the asymptotic value of the MI becomes insensitive of $N$ for $N \gtrapprox 10$, see Fig.~\ref{cEEfig6}(a). The infinite temperature asymptotic value of the MI is understood from Eq.~\eqref{eq. asymptotic and T = inf MI}. Assuming that any of the $2^N$ trajectories can be found in any of the $2^N$ pockets of the phase space with equal likelihood, we have that $p_{\bm{\sigma}} = 2^{-N} x_{\bm{\sigma}}$, with $x_{\bm{\sigma}}$ a random variable with binomial distribution of parameters $2^N$ and $2^{-N}$. This converges, for $N \to \infty$, to a Poisson distribution of parameter $\lambda = 1$, i.e., $x_{\bm{\sigma}} = k$ with probability $\frac{1}{e k!}$. From Eq.~\eqref{eq. asymptotic and T = inf MI}, we thus get that the asymptotic MI at infinite effective temperature is
	\begin{equation}
	I_{A;B}^{\infty} \approx \sum_{k = 1}^{\infty} \frac{\log(k+1)}{e k!} \approx 0.5734,
	\label{eq. asymptotic MI Heisenb}
	\end{equation}
	that perfectly matches the numerics in Fig.~\ref{cEEfig6}(a). By contrast, at finite temperature ($q_0 = 0.7)$ the asymptotic MI is proportional to $N$, i.e., extensive, see Fig.~\ref{cEEfig6}(b). Our numerics thus confirms the analytical prediction in Eq.~\eqref{eq. asymptotic MI}: the asymptotic MI fulfills area- and volume-law scaling at infinite and finite temperature, respectively.
	
	As for the cEE, we instead find that both at finite and infinite temperature the asymptotic value fulfills volume law scaling, see Fig.~\ref{cEEfig6}(c,d). Indeed, extensivity of the cEE is understood not dissimilarly from the quantum case: a wavefunction with randomized components (in the computational basis) leads to an extensive cEE~\cite{page1993average}. In the classical case, the randomness of the classical wavefunction, in Eq.~\eqref{eq. classical wavefunction}, ultimately emerges as a result of chaos and incompressibility, that render the probability distribution highly structured, unlike a thermal one.
	
	The different scaling of the asymptotic value of the MI and the cEE with system size at an infinite effective temperature is a major difference that we find between the two information measures. In fact, in the analogue quantum case, pure states at an infinite effective temperature do result in volume-law scaling asymptotic EE~\cite{page1993average}, which highlights the necessity for the cEE to be introduced as the close classical counterpart of the EE, able to capture features that the classical MI misses.
	
	We perform a more in-depth scaling analysis in Fig.~\ref{cEEfig7}. In Fig.~\ref{cEEfig7}(a) we plot $I_{A;B}(t = \infty)$ and $S_e(t = \infty)$ versus the single-spin initial probability $q_0$. As expected, in the trivial limits $q_0 = 0,1$, corresponding to initially fully polarized spins and no dynamics, the MI and cEE remain zero at all times. For any other values of $q_0$, nontrivial dynamics occurs, information spreads, and the asymptotic MI and cEE take a finite value. However, while volume-law scaling is observed for the cEE at any $0<q_0<1$, the MI is extensive only for $q_0 \neq 0,0.5,1$, whereas it fulfills area-law scaling at infinite temperature for $q_0 = 0.5$. Fig.~\ref{cEEfig7}(b) helps to better appreciate the specific analytic form of the scaling, that is linear in system size $N$. With a linear fit we can extract the scaling coefficient $\alpha$, which we plot in Fig.~\ref{cEEfig7}(c) versus the single-spin initial probability $q_0$, providing a comprehensive confirmation of our assessments concerning the scaling of MI and cEE.
	
	\section{Many-body information in practice -- bridging the procedural gap}
	\label{sec. cl vs qu}
	
	As we have shown, many of the key features of many-body information spreading are remarkably similar in the quantum and classical settings. The two, however, are fundamentally different because the many-body wavefunction $\ket{\psi}$ can be encoded in a physical system, whereas the classical many-body probability distribution cannot \footnote{As discussed in Section~\ref{sec. CA}, the probability distribution is implicitly evolved by the system as long as the state of the system is not observed. This is in analogy with the quantum case, in which the evolution of the wavefunction proceeds undisturbed as long as one does not measure the system. The two cases, however, differ in that the quantum wavefunction is composed of a superposition of states, which is physical rather than reflecting our ignorance on the system.}. However, we argue here that this difference is of limited practical relevance when measuring or computing the EE, cEE, or MI.
	
	To obtain the quantum EE, two routes can be taken. (i) The first is to evolve the wavefunction on a classical computer, and use it to compute the EE. This requires exponentially large computational resources (memory and time). (ii) The second is to evolve the wavefunction on a physical quantum system, e.g., a quantum computer, which does not require exponentially large resources. However, computing the EE requires full tomography of the system and thus exponentially many runs~\cite{nielsen2002quantum}. Furthermore, once reconstructed, the wavefunction should again be stored on a classical computer, where with exponentially large resources one can then compute the EE.
	
	The classical scenario is completely analogous, to the point that we can almost duplicate the previous paragraph: In the classical case, two routes can be taken to compute the dynamics of the cEE. (i) The first is to evolve the probability distribution on a classical computer through Eq.~\eqref{eq. F}, and use it to compute the cEE. This route requires exponentially large computational resources (memory and time). (ii) The second is to run the actual classical dynamics itself on a physical implementation of a cellular automaton. Extracting the many-body probability distribution requires configuration statistics over exponentially many runs. Furthermore, once reconstructed, the probability distribution should again be stored on a computer, where with exponentially large resources one can then compute the cEE.
	
	This remarkably close procedural classical-quantum analogy continues when introducing measurements as in Section~\ref{sec. measurements}, which again requires one of the two routes (i) and (ii) above. In particular, to obtain the EE in an experiment, the following would be necessary. A first run of the experiment would return a list of measurement outcomes $\{m_j(t)\}$ and a physical state encoding the many-body wavefunction. To access the latter and compute the EE, however, one would need to run the experiments exponentially many times, making sure that the final to-be-measured wavefunction is always the same. That is, in the following runs one should not only repeat the measurements at the same spacetime locations, but also make sure that their outcome matches that of the first run, $\{m_j(t)\}$, which requires post-selection. Indeed, this procedure perfectly matches the classical one described in Section~\ref{sec. measurements} and Fig.~\ref{cEEfig3}(c).
	
	\section{Discussion and conclusion}
	\label{sec. discussion} 
	
	In this work, we have shown that many of the celebrated features of quantum many-body information spreading can also appear in a classical setting, provided that one explicitly accounts for the whole exponentially large probability distribution, without making any assumption on its equilibration. Within this framework, we have shown that key phenomenology of quantum EE dynamics, namely linear growth until saturation to an extensive value, also occurs in both the classical MI and the `classical EE' (cEE) defined above. We verified this behaviour for both cellular automata and Hamiltonian systems. For the former, we highlighted time-reversibility to play a key role in information spreading, and found the first instance of a classical MIPT, in direct correspondence with its quantum counterpart. As for Hamiltonian dynamics, we instead provided a simple expression for the asymptotic MI, which yields volume- and area-law scaling for states at an effective finite and infinite temperature, respectively. We verified this prediction numerically in the case of the Heisenberg chain, and showed it is in contrast to the cEE, whose asymptotic value is extensive even at infinite effective temperature, like in the quantum case. Finally, we argued that the study of classical and quantum information spreading requires in practice analogous procedures, involving exponentially large resources in both cases.
	
	Since EE is defined for quantum systems, one may naively assume that effects that are described or measured through it are also purely quantum. Of course, the interplay of carefully engineered interference effects and many-body entanglement can lead to striking quantum effects, for instance enabling quantum computing~\cite{nielsen2002quantum}. By contrast, the entanglement generated in the chaotic dynamics relevant in the context of thermalization is generally `uncontrolled'. Rather than allowing to efficiently perform a computational task, it requires exponentially large resources or many runs for its characterization, effectively resulting in physics that is much more classical than one could have expected. Indeed, here we have shown that a whole range of dynamical behaviours of the EE also emerges in the context of classical correlation spreading.
	
	The idea of sampling and evolving many classical initial conditions adopted here is reminiscent of phase space methods like the Truncated Wigner Approximation (TWA)~\cite{polkovnikov2002nonequilibrium, sinatra2002truncated, polkovnikov2003quantum, polkovnikov2010phase, schachenmayer2015many}. The mission of the latter is to efficiently simulate the dynamics of quantum many-body systems by evolving an ensemble of classical trajectories, for which the sampling of the initial conditions is constructed so to most faithfully mimic quantum fluctuations. A classical-quantum link for EE has also been established for quantum systems with a well-defined classical limit, for which the EE dynamics can be understood in relation to the presence (or lack) of underlying classical chaos~\cite{lerose2020bridging}.
	
	The philosophy of our work contrasts with that mission: while we got inspired by the machinery of quantum mechanics to frame the classical problem on par with the quantum one, our goal was not to construct a classical model that could quantitatively emulate quantum physics, nor to link the classical and quantum descriptions in the semiclassical limit. Instead, we aimed at showing that the main qualitative features of many-body information spreading are analogous in the classical and quantum settings. If the classical-quantum quantitative agreement of few-point observables in the TWA is by construction, in the sense that it requires devising a classical protocol that emulates the quantum one, the classical-quantum qualitative agreement of multi-point information measures in this work is intrinsic, in the sense that it just happens, from the most natural setting of initially independent local fluctuations that become nonlocally correlated through dynamics. In light of this philosophy, our classical treatment did not aim at being a simple and efficient approximation of quantum mechanics, but conversely at showing that classical and quantum information spreading can be equally complex and rich.
	
	Looking forward, our work opens a broad spectrum of research questions. Indeed, within the framework and with the tools that we have outlined here, the study of classical information spreading could be as fruitful as it has been in the quantum case. For instance, a very interesting question regards the role of strong disorder in classical information spreading. In the quantum case, disorder can lead to MBL, which is characterized by a peculiar slow growth $\sim \log t$ of the EE, until reaching asymptotic extensivity~\cite{bardarson2012unbounded}. In the classical case MBL is not possible but transport is nonetheless strongly suppressed~\cite{oganesyan2009energy}, and a slow growth of the cEE is expected. The precise characterization of the functional form of such a growth is certainly worth studying. On a related line, further research should investigate what role the counterparts of integrability~\cite{buvca2021rule,wilkinson2022exact} and conserved quantities~\cite{deger2022arresting} play in classical information spreading. The cEE could then shed new light on classical prethermalization~\cite{rajak2018stability, mori2018floquet, rajak2019characterizations, howell2019asymptotic} and prethermal discrete time crystals~\cite{pizzi2021classicala, pizzi2021classicalb, ye2021floquet}. As well, it is now natural to expect that the research thread on quantum MIPT~\cite{skinner2019measurement,li2019measurement,bao2020theory,choi2020quantum,tang2020measurement,jian2020measurement,zabalo2020critical,tang2020measurement,nahum2021measurement,ippoliti2021entanglement,block2022measurement} would carry over to the classical realm, in which the study of different kinds of automata, initial conditions, and measurement protocols would be worthwhile.
	
	The many-body probabilistic framework that we established could then be used to firmly address the question what the strict classical counterparts of other information-related quantities from current quantum research are. For instance, it could be used to define a classical out-of-time order correlator (OTOC) going beyond the decorrelator in Refs.~\cite{bilitewski2018temperature,liu2021butterfly} (based on just two copies of the system, rather than on a probability distribution over exponentially many of them).
	
	Finally, on a higher and open-ended level, our work opens the way to quantum-inspired advances in physics, information theory, and statistics. Indeed, while quantum EE can only be applied to pure quantum systems, the cEE that we introduced in Eq.~\eqref{eq. cEE} has potentially a very broad range of applicability. Beyond the information spreading considered here, it is suited for \emph{any} probability distribution. Indeed, the cEE can capture some features that the MI misses (e.g., here, extensivity at infinite temperature), and could, thus, prove a powerful quantum-inspired tool to characterize new aspects of classical correlations.
	
	\textbf{Acknowledgements.}
	We thank C.~Castelnovo, S.~Choi, P.~Claeys, A.~G.~Green, M.~Hofstetter, I.~Kaminer, A.~Lamacraft, L.~Pacchiardi, and N.~Y.~Yao for insightful discussions related to this work. We acknowledge support from the Imperial-TUM flagship partnership. A.~P.~acknowledges support from the Royal Society and hospitality at TUM. D.~M.~acknowledges funding from ERC Advanced Grant QUENOCOBA under the EU Horizon 2020 program (Grant Agreement No.~742102). This research is part of the Munich Quantum Valley, which is supported by the Bavarian state government with funds from the Hightech Agenda Bayern Plus.

	\textbf{Author contributions.}
	A.~P. conceived the study, performed the calculations, and drafted the manuscript. All the authors contributed with continued discussion of ideas and results, and with the finalization of the writing.

\end{document}